\documentclass[twocolumn,prd,floatfix,nofootinbib,longbibliography,notitlepage,superscriptaddress]{revtex4-1}

\pdfoutput=1

\usepackage{amsmath, amssymb, amsfonts, amsthm, latexsym, mathrsfs, bbm, slashed}

\usepackage[usenames,dvipsnames]{xcolor}
\usepackage{graphicx}

\usepackage{xifthen}

\usepackage[inline]{enumitem}

\usepackage{setspace}
\usepackage[marginal, multiple]{footmisc}

\usepackage[T1]{fontenc}
\usepackage[utf8]{inputenc}
\usepackage{lmodern}
\usepackage[normalem]{ulem}

\usepackage[unicode, colorlinks, allcolors=blue!70!black, linktocpage, pdfusetitle]{hyperref}
\definecolor{CiteColor}{rgb}{0.55,0,0}
\hypersetup{citecolor=CiteColor}
\definecolor{RefColor}{rgb}{0,0.5,0}
\hypersetup{linkcolor=RefColor}
\usepackage[all]{hypcap}

\usepackage{orcidlink}

\usepackage{verbatim}

\usepackage{cancel}

\usepackage{microtype}

\usepackage{standalone}
\usepackage{tikz}
\usetikzlibrary{arrows.meta}
\usetikzlibrary{decorations.markings}
\ifdefined\myext
\usetikzlibrary{external}
\fi

\newcommand{\nn}{\nonumber}

\newcommand{\pd}{\partial}
\newcommand{\cd}{\nabla}

\newcommand{\eff}{\text{eff}}

\newcommand{\SeffdotL}{\vec{S}_{\eff}\cdot \vec{L}}

\newcommand{\pb}[1]{\left\lbrace  #1   \right\rbrace  }
\newcommand{\avg}[1]{\left\langle #1 \right\rangle}

\newcommand{\figschemesetup}{%
\begin{figure}
  \includegraphics[width=\linewidth]{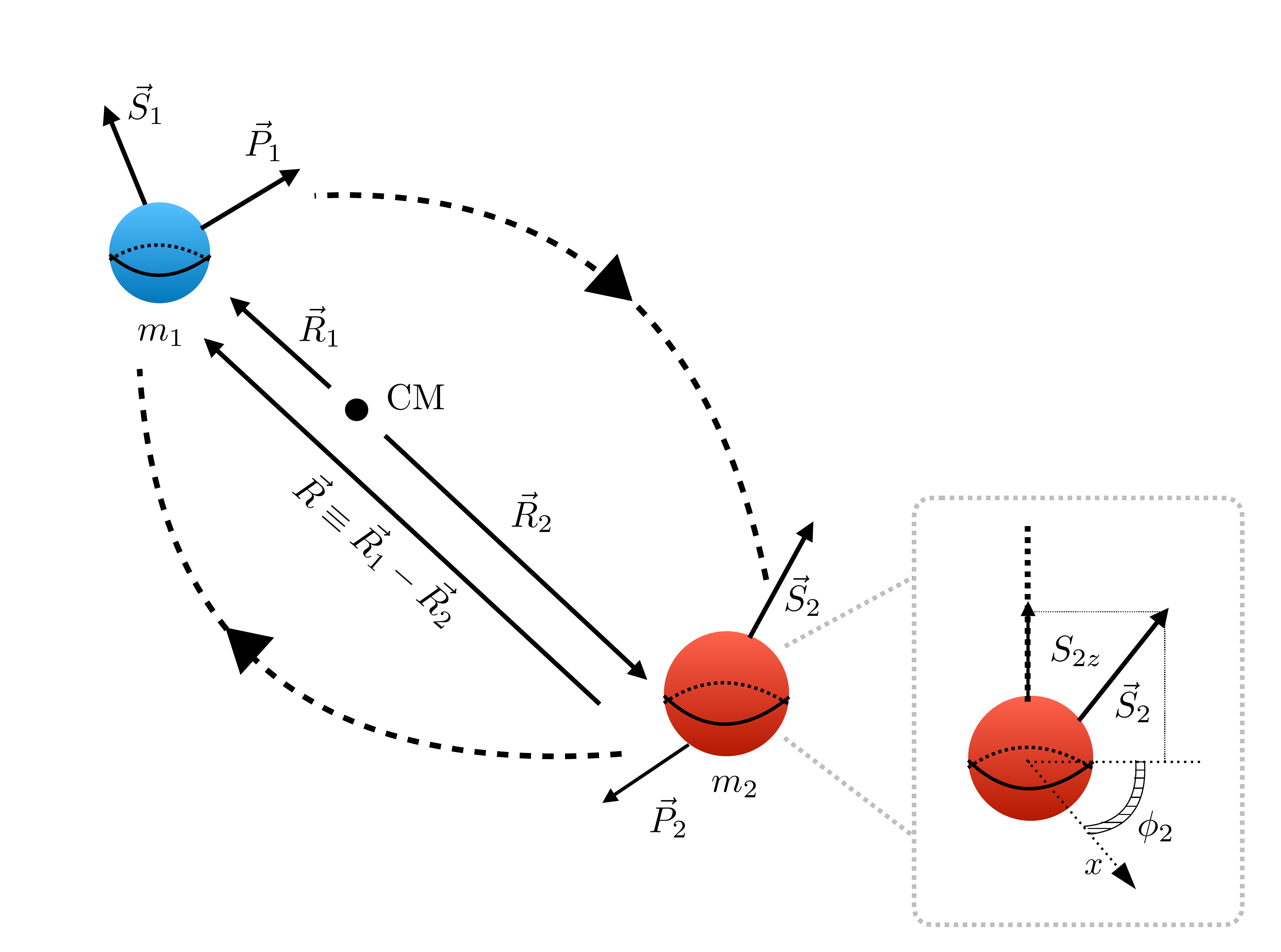}
  \caption{Schematic setup of a precessing black hole
    binary. Positions, velocities and momenta are all defined as
    Newtonian vectors built from the center-of-mass.
    \vspace{-1.5em}}
  \label{fig:SetupFig}
\end{figure}
}

\newcommand{\figloopint}{%
\begin{figure}
  \includegraphics[width=1.05\linewidth]{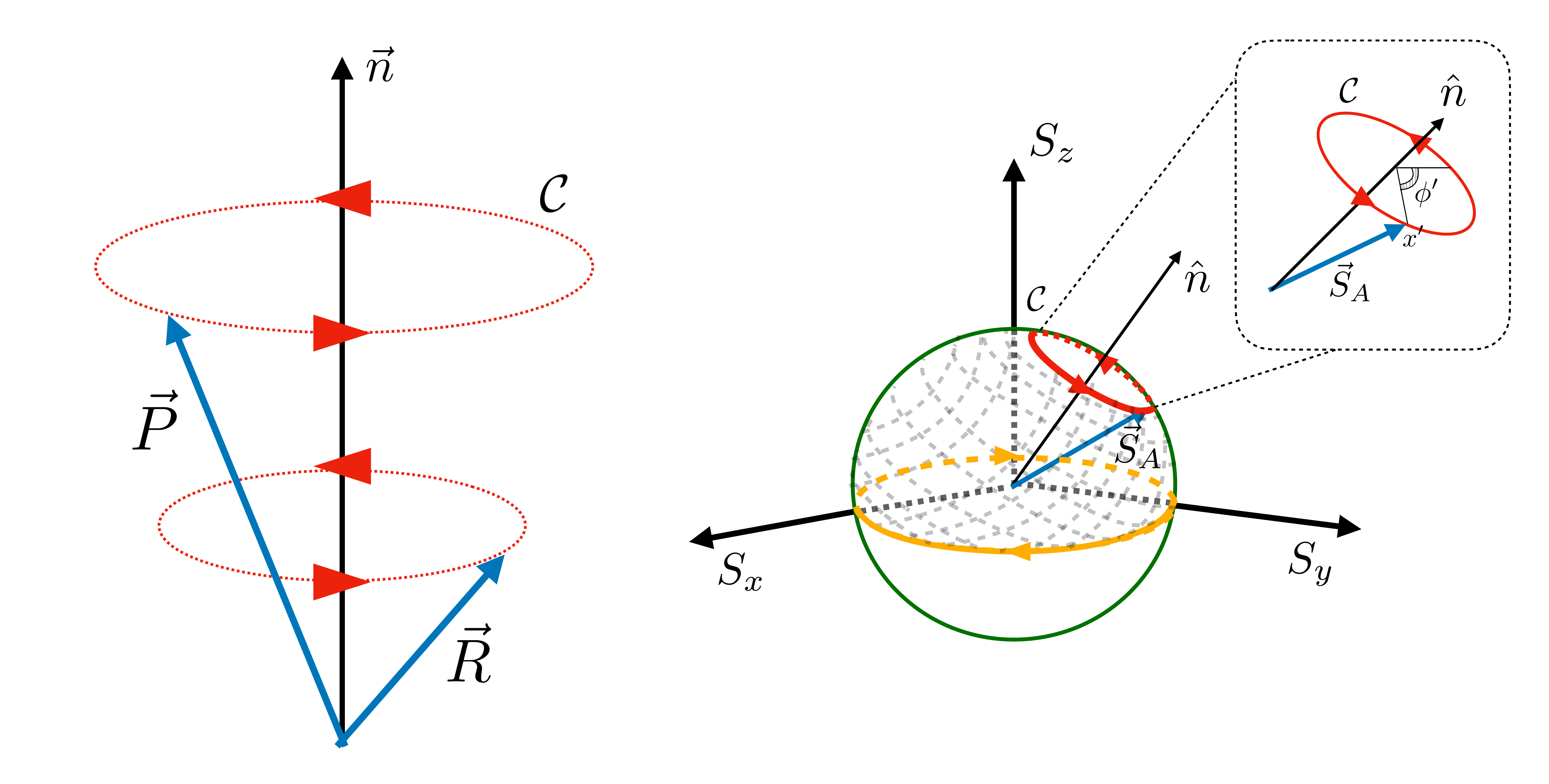}
  \caption{Configuring the integration paths for the action integrals.
     Left panel: Orbital loops corresponding to different equivalence
    classes while having the same topology. Right panel: Spin integration area `capped'
    by the equatorial plane, in orange, and the 3D projection of the loop $\mathcal{C}$ in red.
     The angle $\phi'$ shown in the zoomed patch coincides with the azimuthal angle $\phi$ in
      Fig.~\ref{fig:SetupFig} when $\hat{n}$ is parallel to $\hat{z}$.}
  \label{fig:loop_int}
\end{figure}
}

\begin{document}

\title{Integrability of eccentric, spinning black hole binaries\\ up to second post-Newtonian order}

\newcommand{\UMiss}{\affiliation{Department of Physics and Astronomy, The University of Mississippi, University, MS 38677, USA}}

\author{Sashwat Tanay\,\orcidlink{0000-0002-2964-7102}}
\email{stanay@go.olemiss.edu}
\UMiss
\author{Leo C.~Stein\,\orcidlink{0000-0001-7559-9597}}
\email{lcstein@olemiss.edu}
\UMiss
\author{Jos\'e T.~G\'alvez Ghersi\,\orcidlink{0000-0001-7289-3846}}
\email{jgalvezg@cita.utoronto.ca}
\UMiss
\affiliation{Canadian Institute for Theoretical Astrophysics,
 University of Toronto, 60~St.~George Street, Toronto, ON M5S 3H8, Canada}

\hypersetup{pdfauthor={Tanay, Stein, and G\'alvez Ghersi}}

\begin{abstract}
  Accurate and efficient modeling of the dynamics of binary black
  holes (BBHs) is crucial to their detection and parameter estimation
  through gravitational waves, both with LIGO/Virgo and LISA. 
  General BBH configurations will have misaligned spins and eccentric
  orbits, eccentricity being particularly relevant at early
  times. Modeling these systems is both analytically and numerically
  challenging.  Even though the 1.5 post-Newtonian (PN) order is
  Liouville integrable, numerical work has demonstrated chaos
  at 2PN order, which impedes the existence of an analytic solution.
  In this article we revisit integrability at both 1.5PN and 2PN
  orders. At 1.5PN, we construct four (out of five) action integrals.
  At 2PN, we show that the system is indeed integrable -- but in a
  perturbative sense -- by explicitly constructing five
  mutually-commuting constants of motion. Because of the KAM theorem,
  this is consistent with the past numerical demonstration of chaos.
  Our method extends to higher PN orders, opening the door for a fully
  analytical solution to the generic eccentric, spinning BBH problem.
\end{abstract}

\maketitle

\section{Introduction}
To date, Advanced LIGO and Virgo have confidently detected 50
gravitational-wave events~\cite{Abbott:2016blz,
  LIGOScientific:2018mvr, Abbott:2020niy}, all of them from compact
binary mergers.  Of these, at least 46 are due to a binary black hole
(BBH) system.  Both detecting and characterizing these systems relies
on computing accurate and efficient waveform templates.
Present waveform models~\cite{Ossokine:2020kjp, Khan:2018fmp,
  Pratten:2020ceb} are already rather sophisticated, including modeling
precession due to spin-orbit coupling; but typically, the orbital
motion is modeled as quasi-circular, and the precession is approximate
(except for NR surrogates~\cite{Varma:2019csw}).
The fact that most eccentricity
should be radiated away by the time of merger has been long
known~\cite{Peters:1963ux,Peters:1964zz}.
Despite constraints on eccentricity~\cite{Romero-Shaw:2019itr}, there
have been tentative claims that some LIGO events were highly
eccentric~\cite{Gayathri:2020coq}.
Moreover with the LISA
mission~\cite{Cutler:1997ta,Audley:2017drz} on the horizon,
eccentricity is expected to play a more prominent
role~\cite{Roedig:2011rn, Chen:2017gfm, Fang:2019dnh}, and may be
especially important for multi-band systems~\cite{Sesana:2016ljz}.

This brings us to the challenge of modeling ``generic'' BBH systems:
two BHs, with their spins misaligned from the orbital angular
momentum, in an eccentric orbit. Eccentricity leads to apsidal
precession, and spin-orbit coupling leads to precession of both the
spins and the orbital plane. Such complicated nonlinear dynamics in a
high dimensional phase space leads to the fear of chaos.
One ultimate goal of studying the BBH problem is to produce rapid
gravitational-wave predictions---and chaos would obstruct the
possibility of analytical waveforms.
Showing the integrability of the system and the existence of
action-angle variables opens the door to constructing a closed-form
analytical waveform model.

The study of chaos and integrability in the spinning, eccentric BBH system has an
interesting history~\cite{Levin:1999zx, Levin:2000md, Hughes:2001gh,
  Schnittman:2001mz, Cornish:2003ig, Schnittman:2004vq, Hartl:2004xr,
  Konigsdorffer:2005sc, Wu:2010mv, 2011GReGr..43.2185W, Mei:2013uqa,
  Huang:2014ska, Wu:2015cqa, Huang:2016vfk}. We will recap some of the
highlights below. Some of the claims in the literature seem at odds
with each other. Besides our main results, we will also explain these
apparent contradictions and correct some misstatements in the
literature regarding integrability of the BBH system.

The generic BBH system, in Hamiltonian form, has long been known to be
integrable at the 1.5 post-Newtonian (PN) order~\cite{Damour:2001tu}.  This
comes from the Liouville-Arnold theorem~\cite{arnold, jose}: the
ten-dimensional phase space has five independent constants of motion,
which all pairwise commute under the Poisson bracket.
This integrability leads to the existence of an analytic
solution~\cite{Cho:2019brd}.
At 2PN, Levin~\cite{Levin:1999zx, Levin:2000md} performed numerical
simulations and concluded that the generic BBH system is
chaotic. Schnittman and Rasio~\cite{Schnittman:2001mz} also simulated
generic systems at 2PN, and by measuring the Lyapunov exponent, found
either no chaos, or weak chaos with a Lyapunov time which was many
times greater than the inspiral time.  Soon after, Cornish and
Levin~\cite{Cornish:2003ig} found the Lyapunov and inspiral time-scale
could be comparable to each other, though they warned that the
Lyapunov time is coordinate-dependent.  Hartl and
Buonanno~\cite{Hartl:2004xr} performed a survey of generic orbits,
simulating them at 2PN (and including some PN terms that previous
authors had not).  For the most part, they found regular (i.e.\
non-chaotic) orbits, though they did report chaos in some cases, which
they reported to be astrophysically disfavored.
Though not discussed in any of these works, the coexistence of regular
and chaotic orbits in phase space is a typical characteristic of a
nearly-integrable system, proven in the KAM theorem~\cite{arnold,
  jose}. This applies to the second and higher PN Hamiltonians, when
treated as a perturbation to the integrable 1.5PN Hamiltonian.

There have also been a number of analytical studies of integrability.
Damour~\cite{Damour:2001tu} pointed out the additional constants of
motion, though did not emphasize that they commute or that the generic
BBH is integrable. Königsdörffer and
Gopakumar~\cite{Konigsdorffer:2005sc, Gopakumar:2005zz} suggested
integrability at higher PN order, by constructing an analytic solution
for two specific mass/spin configurations, and removing all spin terms
in the Hamiltonian except for the leading order spin-orbit
interaction.  Beyond
the 1.5PN spin-orbit effect, the next non-trivial effect on
integrability is the spin-spin interactions at 2PN, which is
conjectured to source chaotic behavior~\cite{Levin:1999zx}.
Let us also mention that some analytic work~\cite{Wu:2010mv,
  2011GReGr..43.2185W, Mei:2013uqa, Huang:2014ska, Wu:2015cqa,
  Huang:2016vfk} has discussed integrability by only counting the
number of constants of motion, which is not enough for the
Liouville-Arnold theorem: the constants must be mutually
commuting. For example, while each of the three components $J_{i}$ are
constants, they do not commute with each other.

Along independent lines, a large body of literature has been developed
by taking advantage of orbit-averaging and precession-averaging. The
principle at work is that there is a large separation of timescales,
$t_{\text{orb}} \ll t_{\text{prec}} \ll t_{\text{rad}}$; so the
orbital variables' influence on precession dynamics may be
approximated by averaging, and similarly for precession-averaging.
Early post-Newtonian works invoking orbit-averaging to study spin
effects include~\cite{Kidder:1995zr, Schnittman:2004vq,
  Racine:2008qv}, and precession-averaging followed
in~\cite{Kesden:2014sla, Gerosa:2015tea}.  An important milestone was
Racine's discovery that a quantity $\vec{L} \cdot \vec{S}_{0}$ (to be
introduced later) is constant under the \emph{Newtonian-orbit-average}
of the 2PN equations of motion (EOMs), despite not being constant under the
full 2PN equations.  We will briefly comment on the relation of our
results to the averaged results.

In this paper, we study the problem of integrability at two levels:
we find the action variables at 1.5PN, and show integrability at 2PN.
These are both part of the larger program to eventually build
analytical waveform models for the generic spinning, eccentric BBH
system.
The known integrability at 1.5PN implies the existence
of action-angle variables.  We derive four (out of the five) action
variables, with the fourth one being in the form of a PN series.
These action variables are closely related to the Keplerian-like
parameterization for the generic system at 1.5PN recently presented in
Ref.~\cite{Cho:2019brd} (that work omitted the 1PN orbital terms from
the Hamiltonian for simplicity, but the approach will work with the
1PN terms included).
We then proceed to 2PN, where in the spirit of perturbation
theory, we add an ansatz for PN corrections to the 1.5PN
exactly-commuting constants, and solve for these corrections to find
the 2PN-valid constants. We work with the full 2PN Hamiltonian rather
than removing the spin-spin interaction. This shows (via the
Liouville-Arnold theorem to be discussed later) that the generic
BBH is integrable at 2PN, in the sense of perturbation theory.
That is, these 2PN constants only mutually commute up to
sufficiently-high-order errors.
This also implies that the action variables can be pushed to 2PN, so
an analytical orbital solution is possible at this order.
We finally revisit the criteria for
integrability by analyzing the timescales for ``constants'' to vary
when evolved with the next order Hamiltonian. With this more physical
criterion, $\SeffdotL$ actually varies at the 1PN timescale, despite
being a 1.5PN constant of motion. The 2PN constants we construct only
vary at 2.5PN order, justifying that the BBH system is integrable at
2PN order.

The existence of these perturbative constants is not in conflict with
the presence of chaos in phase space.  From the KAM theorem, most
invariant tori will remain unbroken under a sufficiently small
perturbation.  Resonant tori will be the first to break up into
chaotic regions.  Our constants are applicable to unbroken tori, which
according to Ref. \cite{Hartl:2004xr} fill the vast majority of phase
space.

The layout of this paper is as follows.
In Sec.~\ref{sec:setup} we introduce preliminaries like post-Newtonian
power counting, Liouville integrability, the Hamiltonian phase space
and Poisson bracket structure for the BBH problem, and the 2PN
Hamiltonian.
In Sec.~\ref{sec:act_1.5PN}, we compute four out of five action
variables up to 1.5 PN by integrating along closed loops on the
invariant tori in phase space.
In Sec.~\ref{sec:integrability_2PN}, we give an algebraic definition
of PN involution and integrability. We then describe how to
systematically construct appropriate ansätze for corrections to add to
constants of motion, reducing the problem to linear algebra. Finally
we solve for the corrections and present the five approximate
constants of motion, which are in involution up to errors that can be
ignored at 2PN.
In Sec.~\ref{sec:discussion}, we present our discussion, ideas for
future work, and conclude.

\section{The setup}
\label{sec:setup}
\figschemesetup

We start by describing the canonical variables and the dynamical
setup used to study eccentric binaries of black holes with precessing
spins in the PN approximation. The BBH system under consideration is
schematically displayed in Fig.~\ref{fig:SetupFig}, using its
center-of-mass frame \cite{Damour:1988mr}, to define the separation
vector $\vec{R} \equiv \vec{R}_1 -\vec {R}_2$ and the linear momenta
$\vec{P} \equiv \vec{P}_1 = - \vec{P}_2$ of a binary of black holes
with masses $m_1$ and $m_2$. With these quantities, we build the
Newtonian orbital angular momentum $ \vec{L} \equiv \vec{R} \times \vec{P }$,
and the total angular momentum
$\vec{J} \equiv \vec{L} + \vec{S}_{1}+ \vec{S}_{2}$ which includes the BH
spins $\vec{S}_{1}$ and $ \vec{S}_{2}$. The individual BH masses are
$m_1$ and $m_2$ and the total mass $M \equiv m_1+m_2 $.
Additionally, the reduced mass is given by $\mu\equiv m_1m_2/M$ and the symmetric
mass ratio $\nu\equiv\mu/M$ is a function of the reduced mass.
The constants $\sigma_1 \equiv 1 + 3 m_{2}/4 m_{1}$ and
$\sigma_2 \equiv 1 + 3 m_{1}/4 m_{2}$ are used to build the effective
spin
\begin{equation}
  \vec{S}_{\mathrm{eff}} \equiv  \sigma_1 \vec{S}_{1} + \sigma_2\vec{S}_2 \,.
  \label{eq:seff}
\end{equation}
This should not be confused with other common spin parameters used in
the literature~\cite{Damour:2001tu, Racine:2008qv, Ajith:2009bn},
namely the projected effective spin
$\chi_{\eff} \equiv (m_1 \chi_1 + m_2 \chi_2)/M$, or the combination
$ \vec{S}_0  \equiv (1+m_2/m_1)   \vec{S}_1 +  (1+m_1/m_2) \vec{S}_2 $.
Racine found~\cite{Racine:2008qv} that $\vec{L}\cdot \vec{S}_{0}$ is
conserved under the Newtonian-orbit-average of the 2PN equations of
motion; we will discuss this further in
Sec.~\ref{sec:integrability_2PN}.

Even when our approach throughout this paper
is purely Hamiltonian, we may define a velocity $\vec{v} \equiv \vec{P}/\mu$ 
since the ratio $v^2/c^2$ is often used as a PN expansion 
parameter. Latin indices $i=1,2,3$ denote the $i$th Cartesian component of a vector, and we
employ the Einstein summation convention unless stated otherwise.

The spin angular momentum for a Kerr black hole labeled $A$ is
\begin{align}
\vec{S}_{A} = \vec{\chi}_{A} \frac{G m_{A}^2}{c}
\,,
\label{eq:pnc17}
\end{align}
where $|\vec{\chi}|\le 1$ so that there are no naked singularities.  Notice
the factor of $1/c$, which affects the post-Newtonian order of any
terms containing spins; this will be detailed in
Secs.~\ref{sec:SecPNC} and \ref{sec:post-newt-hamilt}.

\subsection{Counting post-Newtonian orders}
\label{sec:SecPNC}

Post-Newtonian counting applies to any function $y$ of phase-space variables, which
we expand as an asymptotic series using a certain PN parameter $x$,
i.e. $y = \sum_{k} y_{k} x^{k}$.  Depending on context, one of $v$, an
orbital frequency $\omega$, or $R$ is used as the expansion parameter.
Specifically, from Newtonian order, we may define $x$ to be any of
\begin{align}
  x \equiv \frac{v^2}{c^2},\quad \left(  \frac{G M \omega}{c^3} \right)^{2/3}, \quad\text{or} \quad \frac{G M}{c^2  R}
  \,.
  \label{eq:pnc12b}
\end{align}
Since we have kept the powers of $c$ explicitly, we can see that any
choice is equivalent to counting powers of $c^{-2}$.  This latter
observation is important when spins are involved, since spin includes
$1/c$ [see Eq.~\eqref{eq:pnc17}] but does not scale with $v, \omega$, or
$R$.

Let a phase-space function $y$ be written in the form
\begin{align}
  y = x^m\sum_{k=0}^{\infty} Y_k x^k
  \,,
  \label{eq:pnc13}
\end{align}
In Eq.~\eqref{eq:pnc13}, $Y_0 \neq 0$ is the first non-vanishing term
in the expansion, and
we would say that the term $Y_{k}$ is $k$PN orders higher than
$Y_{0}$, or is of ``relative $k$PN order.''  For example, when
including spins in the total angular momentum,
\begin{equation}
  \vec{J} = \vec{L} + \vec{S}_{1} + \vec{S}_{2} = \vec{L} \left[1+\mathcal{O}\left(\frac{v}{c}\right)\right]
  \,,
  \label{eq:pnc19}
\end{equation}
we see that spins are 0.5PN orders higher than orbital angular
momentum.

\subsection{Hamiltonian dynamics on a symplectic manifold}
\label{sec:Hamilton}

From now on, we will follow the Hamiltonian formulation to study the
BBH system; we will shortly review its algebraic structure \cite{jose,
  arnold}. Hamiltonian dynamics takes place on an (even-dimensional)
symplectic manifold.  A smooth manifold equipped with a closed
non-degenerate differential 2-form $\Omega$ (the symplectic form) is
called a symplectic manifold. The algebra of non-vanishing Poisson
brackets (PBs) between the phase-space variables $R^i,P_{j},S_1^i$, and
$S_2^i$ is given by
\begin{align}
  \left\lbrace  R^i, P_j   \right\rbrace = \delta^{i}_j
  \quad\text{and}\quad
  \left\lbrace S_{A}^i, S_{B}^j  \right\rbrace  = \delta_{AB}\epsilon^{ij}{}_{k} S_{A}^k
\,.
\label{eq:shd2}
\end{align}
Notice that all brackets with spins preserve the norms $|\vec{S}_{A}|$, so
although the spin vectors are three-dimensional, each is restricted to
evolve on the surface of a 2-sphere.  This makes the phase space a
ten-dimensional manifold.

Time evolution under a Hamiltonian $H$ of any phase-space quantity
$f(\mathcal{Q}^i,\mathcal{P}_{i})$ is given by
$\dot{f} = \left\lbrace f,H \right\rbrace$, where $\mathcal{Q}^i,\mathcal{P}_{i}$
collectively denote canonical coordinates on phase space. The standard rules of
sum, product, anti-commutativity, and chain rule make the PBs in
Eq.~\eqref{eq:shd2} sufficient to evaluate the PB of any quantities
built from $\mathcal{Q}^i,\mathcal{P}_{i}$.\footnote{%
If computing PBs by hand, the following derived identities are also useful:
$\pb{L^i, L^j} = \epsilon^{ij}{}_{k} L^k$;
and, for any scalar function $f$,
$\pb{f, \vec{L}} = \vec{P} \times \cd_P f + \vec{R} \times \cd_R f$,
where the 3-vector $\cd_{P} f$ has components $\pd f/\pd P^{i}$, and
similarly for $\cd_{R}f$.
}
The remainder of this section is for readers interested in the symplectic
structure, relevant to computing action-angle variables, the subject
of Sec.~\ref{sec:act_1.5PN}.

Our symplectic manifold is the product of the 6-dimensional phase
space of orbital dynamics, and two 2-dimensional spin phase spaces,
each of which is an $S^{2}$ (the only $S^{n}$ that admits a symplectic
structure).  The symplectic form is correspondingly a sum over the
three manifolds.  Commonly, symplectic forms are presented in Darboux
coordinates,
\begin{align}
  \label{eq:OmegaDarboux}
  \Omega & \equiv \sum_i d  {\cal{P}}_i \wedge d {\cal{Q}}^i
  \,.
\end{align}
This is possible on the orbital phase space, which is a cotangent
space, $T^{*}\mathbb{R}^{3}$, and admits the globally-valid canonical
form $\Omega^{\text{orb}} = dP_{i}\wedge dR^{i}$.

However, there is no globally-valid Darboux coordinate system on the
2-sphere.  The symplectic structure on the $S^{2}$ is unique up to
scaling and is proportional to the standard area element,
$\Omega^{\text{spin}}_{ij} \propto \epsilon_{ij}$; the normalization
is fixed to agree with Eq.~\eqref{eq:shd2}.  Thinking of the $S^{2}$
as an embedded submanifold in spin space, the inverse symplectic
form can be written as
\begin{align}
  \label{eq:omega-spin-so3}
  (\Omega_{\text{spin}}^{-1})^{ij} = S^{k}\epsilon_{k}{}^{ij}
  \,.
\end{align}
This
representation should make it clear that the symplectic form is
SO(3) covariant.  An equivalent representation is
$\Omega^{\text{spin}} = dS_{z} \wedge d{\phi}$, where $\phi$ is the
azimuthal angle of the spin about the $z$ axis.
The total symplectic form is thus
\begin{align}
\Omega  =   dP_i \wedge dR^i  +   dS_{1z} \wedge d{\phi}_{1} + dS_{2z} \wedge d{\phi}_{2}
\,.
\label{eq:shd14a}
\end{align}
As noted above, it is SO(3) covariant, which will be useful in
evaluating some action integrals.
Finally let us note that while $\Omega^{\text{orb}}$ is $c$-independent,
while
$\Omega^{-1}_{\text{spin}}$ carries one power of spin [seen in
Eqs.~\eqref{eq:shd2} and \eqref{eq:omega-spin-so3}],
and spin carries a power of $1/c$.  Orbital and
spin PBs thus change PN orders in different ways, which will be
important in Sec.~\ref{sec:integrability_2PN}.

\subsection{Integrable systems}
\label{sec:CIS}
A $2n$-dimensional Hamiltonian system is said to be integrable in the
Liouville sense if there exist $n$ independent phase-space functions
$F_{i}$ which are all mutually Poisson commuting, $\pb{F_{i},F_{j}}=0$.
These functions are said to be ``in involution''~\cite{fasano, arnold, jose}.\footnote{%
  More precisely, the Liouville-Arnold theorem states that, on a $2n$-dimensional symplectic manifold, if
  $\partial_t H=0$ and there are $n$ independent phase-space functions $F_{i}$ in
  mutual involution, and if level sets of these functions form a compact
  and connected manifold, then the system is integrable.
}
Bound systems that are integrable admit a canonical transformation to
a set of phase-space coordinates called action-angle
variables.  The evolution of such systems is trivial in
action-angle variables, so there cannot be any chaos or phase space
mixing; all bound orbits are multiply-periodic.  Action-angle
variables are ideal for studying perturbations of integrable systems.
For our purposes, we would like to treat terms of higher PN orders as a
perturbation of an integrable system.

A level set of all the constants of motion must be an $n$-dimensional
torus $T^{n}$~\cite{arnold}.  The actions $\mathcal{J}_{i}$ can be found via
certain coordinate-independent integrals along $n$ closed loops
restricted to the tori (holding constant each of the $F_{i}$).  If
global Darboux coordinates are possible, the action integrals
are~\cite{fasano, arnold,jose},
\begin{align}
  \mathcal{J}_k = \frac{1}{2 \pi} \oint_{\mathcal{C}_k}   \sum_i  {\cal{P}}_i d {\cal{Q}}^i
  \,.
\label{eq:shd16}
\end{align}
Here $\mathcal{C}_{k}$ is the $k$th loop on the torus.  The set of $n$
loops must be in different homotopy classes (more precisely, the
homotopy classes form an integer lattice $\mathbb{Z}^{n}$, and our $n$
loops' homotopy classes must span the lattice).  The 1-form integrand
of Eq.~\eqref{eq:shd16} is a symplectic potential, $\theta = \sum_{i}\mathcal{P}_{i}d\mathcal{Q}^{i}$, whose exterior
derivative gives the symplectic 2-form,
$\Omega = d\theta$.  Since $\Omega$ is closed, it is
straightforward to show that the $\mathcal{J}_{k}$ depend only on homotopy
class, and not on choice of loop in that class.

However on some symplectic manifolds, including the 2-sphere, $\Omega$
is not an exact form, $\Omega \neq d\theta$.  This makes the action
integrals Eq.~\eqref{eq:shd16} ambiguous.  One approach is to make a
global choice of how to `cap' the loops to another reference loop, and
thus perform integrals of $\Omega$ over 2-surfaces.  This ambiguity is
benign, as it will only shift the action integrals by global
constants.\footnote{We thank Samuel Lisi for discussion of the finer
  points of this ambiguity.}

To complete the coordinate system, there will be $n$ angle variables
$\phi_{i}$ which are conjugate, i.e.\
$\pb{\phi_{i}, \mathcal{J}_{j}} = \delta_{ij}$ and all other PBs vanishing.
Each angle variable $\phi_{i}$ runs from 0 to $2\pi$ as one follows
the flow $d  /d \phi_{i} =  \pb{-,\mathcal{J}_{i}}$
generated by its conjugate action.  We will
not construct the angle variables in this work.

\subsection{2PN Hamiltonian with spins included}
\label{sec:post-newt-hamilt}

To write the Hamiltonian at different post-Newtonian orders, we adopt
the convention that $H_{n\mathrm{PN}}$ stands for the part of
Hamiltonian which is of $n$PN order relative to the leading Newtonian
order term (dubbed $H_{\mathrm{N}}$).  The Hamiltonian up to
2PN of the BBH system in the center-of-mass frame is
\begin{align}
  H  =  H_{\mathrm{N}}   +    H_{1\mathrm{PN}}  +    H_{1.5\mathrm{PN}}   +    H_{2\mathrm{PN}} + \mathcal{O}(c^{-5})
  \,,
  \label{eq:int1}
\end{align}
where $\mathcal{O}(c^{-5})$ represents corrections of order
2.5PN and higher.
To simplify we will use the scaled quantities $\vec{r} \equiv
\vec{R}/GM$, $\vec{p}\equiv \vec{P}/\mu$, and the radial component of
scaled momentum is $\hat{r}\cdot \vec{p}$, with the
implicit understanding that the ``hatted'' version of any
vector in this paper is the corresponding unit vector.
The vector $\vec{p}$ has
units of velocity, and $1/r$ has units of velocity squared, enabling
the easy reading of PN orders.
The individual contributions are~\cite{Barker:1966zz, Damour:2001tu,
  Barker:1975ae, Hartl:2004xr,Steinhoff:2010zz}
\begingroup
\allowdisplaybreaks
\begin{align}
H_{\mathrm{N}}       ={} &   \mu\left(\frac{p^2}{2}     -    \frac{1}{r}\right)  ,      \label{eq:int1a}    \\
H_{\mathrm{1PN}}     ={} &   \frac{\mu}{c^2}\bigg\{  \frac{1}{8}(3 \nu-1) p^4     +  \frac{1}{2r^2}       \nonumber  \\
   &   - \frac{1}{2 r}  \left[   (3+ \nu) p^2  +  \nu (\hat{r}\cdot \vec{p})^2  \right]\bigg\}    ,    \label{eq:int1b}     \\
H_{\mathrm{1.5PN}}   ={} &   \frac{ 2 G }{c^{2} R^{3}} \SeffdotL   ,          \label{eq:int1c}    \displaybreak[0]\\
H_{\mathrm{2PN}}     ={} &   \frac{\mu}{c^4}  \bigg\{
-\frac{1}{4r^3}(1+ 3 \nu) +   \frac{1}{16}\left(1 -5 \nu + 5 \nu ^2  \right) p^6  \nonumber  \\
   &  +  \frac{1}{2r^2} \left(3 \nu  {(\hat{r} \cdot \vec{p})}^2 + (5 + 8 \nu) p^2\right)       \nonumber\\
   &  +  \frac{1}{8r} \bigg[ - 3 \nu ^2   {(\hat{r} \cdot \vec{p})}^4  - 2 \nu ^2    {(\hat{r} \cdot \vec{p})}^2 p^2     \nonumber\\
   &  +  \left( 5 - 20 \nu  - 3 \nu ^2\right) p^4      \bigg]        \bigg\}
   +  H_{\text{SS,2PN}}
   \,.
\label{eq:int1d}
\end{align}
\endgroup
The 2PN spin-spin interaction is
\begin{align}
  H_{\text{SS,2PN}}  & =    H_{\text{S1S1}}  +   H_{\text{S2S2}}   + H_{\text{S1S2}}   ,       \label{eq:int1e} \\
  H_{\text{S1S1}}    & =    \frac{G}{c^{2}}\frac{m_2}{2m_{1}} S_{1}^{i}S_{1}^{j} \ \pd_{i}\pd_{j} R^{-1} \,, \\
  H_{\text{S2S2}}    & =    \frac{G}{c^{2}}\frac{m_1}{2m_{2}} S_{2}^{i}S_{2}^{j} \ \pd_{i}\pd_{j} R^{-1} \,, \\
  H_{\text{S1S2}}    & =    \frac{G}{c^{2}}  S_{1}^{i}S_{2}^{j} \ \pd_{i}\pd_{j} R^{-1} \,,
  \label{eq:int2}
\end{align}
where $\pd_{i}\pd_{j} R^{-1} = (3 \hat{R}_{i}\hat{R}_{j} -
\delta_{ij})/R^{3}$ is symmetric and trace-free.

Notice that since $H_{\text{1.5PN}} \sim \mathcal{O}(c^{-2} S)$ and, as
previously mentioned, spin goes as $S \sim \mathcal{O}(c^{-1})$, so
indeed $H_{1.5PN} \sim \mathcal{O}(c^{-3})$.  Likewise,
$H_{\text{SS,2PN}} \sim \mathcal{O}(c^{-2} S^{2}) \sim \mathcal{O}(c^{-4})$,
justifying the claimed PN orders of these terms. 

\section{Action variables at 1.5PN order}
\label{sec:act_1.5PN}
To start, we will focus on integrability at 1.5PN,
truncating the Hamiltonian to
\begin{align}
  H  =  H_{\mathrm{N}}   +    H_{1\mathrm{PN}}  +    H_{1.5\mathrm{PN}}   + \mathcal{O}(c^{-4})
  \,.
  \label{eq:H1.5PN}
\end{align}
As has been known for many years now~\cite{Damour:2001tu}, truncating at
this order gives a 10-dimensional phase space with 5 constants of
motion $F_{i}$ in mutual involution, namely, the set
$\{F_{i}\} = \{H, J^2, J_z, L^{2}, \SeffdotL\}$.
At this level, the involution is ``exact,'' for the associated PBs vanish exactly.
This involution can be verified by the \textsc{Mathematica} notebook
which accompanies this article~\cite{MmaSupplement}, which makes use
of the \textsc{xAct/xTensor} suite~\cite{JMM:xAct,
  MARTINGARCIA2008597}.

This involution implies the existence of action-angle variables.  We
will construct four out of five action variables in this section.  For
each action variable $\mathcal{J}_{k}$, we will consider a different
loop $\mathcal{C}_{k}$ tangent to the five-torus given by constancy of
the five $F_{i}$, and perform the (capped) loop integral of
Eq.~\eqref{eq:shd16}.

\subsection{Loops generated by $J^2, J_z$, and $L^2$}
\label{sec:flow-loops}

We find three of these loops by following the flow of the generators
$J^{2}, J_{z}$, and $L^{2}$.  To demonstrate, let
$d/d\lambda_1 = \pb{- , L^{2}}$ be the vector field tangent to the flow
generated by $L^{2}$.  Notice that this flow makes $\vec{R}$ and
$\vec{P}$ rigidly rotate about the constant $\hat{L}$, while the two $\vec{S}_{A}$
are not moved. Thus we have (with $\vec{V}$ representing either $\vec{R}$ or $\vec{P}$)
\begin{align}
  \frac{d\vec{V}}{d\lambda_1} &= \pb{\vec{V}, L^{2}} = 2\vec{L}\times\vec{V} \,,
  &
  \frac{d\vec{S}_{A}}{d\lambda_1} &= 0
  \,.
\end{align}
As this is a rigid rotation, the phase-space flow will complete one
cycle as the parameter $\lambda_{1}$ increases by $\Delta \lambda_1 = 2\pi/|2\vec{L}|$.
Similarly, let $d/d\lambda_{2} \equiv \pb{-, J_{z}}$.  This time all vectors
rotate rigidly about the $\hat{z}$ axis,
\begin{align}
  \frac{d\vec{V}}{d\lambda_2} &= \hat{z}\times\vec{V}  ,
\end{align}
with $\vec{V}$ representing any of $\vec{R},~\vec{P},$ and $\vec{S}_A$.
After $\lambda_{2}$ increases by $\Delta \lambda_2 = 2\pi$, the spin and orbital
phase-space variables will close the loop.
Thirdly, with $d/d\lambda_{3} \equiv \pb{-, J^{2}}$, all vectors
rigidly rotate around the constant $\hat{J}$,
\begin{align}
  \frac{d\vec{V}}{d\lambda_3} &= 2\vec{J}\times\vec{V} \,,
\end{align}
with $\vec{V}$ again representing any of $\vec{R},~\vec{P},$ and $\vec{S}_A$.
The phase-space flow under $d/d\lambda_{3}$
closes after the parameter $\lambda_{3}$ increases by $\Delta \lambda_3
= 2\pi/|2\vec{J}|$.

All three of these flows can be treated with the same method.  Since
the symplectic forms on orbital and spin phase spaces simply add, we
treat the orbital and spin components one at a time and add the final
results,
\begin{align}
  \label{eq:Jsum}
  \mathcal{J} &= \mathcal{J}^{\text{orb}} + \mathcal{J}^{\text{spin}} \,, \\
  \mathcal{J}^{\text{orb}} &\equiv \frac{1}{2\pi} \oint_{\mathcal{C}} \sum_{i} P_{i} dR^{i} \,,
\end{align}
and similarly for the spin sector, except that the spin integral is
`capped' so as to become an area integral of $\Omega^{\text{spin}}$.

\figloopint

We write $d/d\lambda$ for any of the three flows, and use $\vec{n}$ to
denote the fixed vector about which others rotate, $\vec{n}$ being one
of $\{2\vec{L}, \hat{z}, 2\vec{J}\}$.  The loop closes after the parameter change of
$\Delta \lambda=2\pi/|\vec{n}|$.  This is illustrated in
Fig.~\ref{fig:loop_int}.  The only exception is that the spin vectors
are not moved by $d/d\lambda_{1}$, but since we break the action
integral up as in Eq.~\eqref{eq:Jsum}, this is simple to implement.
First, when we parameterize $\mathcal{C}$ using $\lambda$, the
$\mathcal{J}^{\text{orb}}$ integral becomes
\begin{align}
  \mathcal{J}^{\text{orb}} &= \frac{1}{2\pi} \int_{0}^{\Delta\lambda} P_{i} \frac{dR^{i}}{d\lambda} \ d\lambda
  = \frac{1}{2\pi} \int_{0}^{\Delta\lambda} \vec{P} \cdot ( \vec{n} \times \vec{R} ) \ d\lambda     \nn  \\
  &= \frac{1}{2\pi} \int_{0}^{\Delta\lambda} \vec{n} \cdot \vec{L} \ d\lambda
  \quad
  = \hat{n} \cdot \vec{L}
  \,.
\end{align}
The second equality comes from evaluating the flow for
$dR^{i}/d\lambda$; the third equality comes from permuting the triple
product.  The last equality arises since in all three cases, $\vec{L}$
rigidly rotates around $\vec{n}$ (because $\vec{R}$ and $\vec{P}$ also rigidly
rotate around $\vec{n}$), so the dot product is constant
around the loop.

For the spin sector, we choose to cap each curve $\mathcal{C}$ by
the equatorial plane (in spin space), i.e.\ the oriented area integral will be
bounded between the $S_{A}^{z}=0$ plane and $\mathcal{C}$. One
can show that this gives the same result as the ordinary integral (for
one of the two spins)
\begin{align}
  \mathcal{J}_{A}^{\text{spin}} = \frac{1}{2\pi} \oint S_{A}^{z} \ d\phi_{A} \,.
\end{align}
While this integral does not seem to be SO(3) covariant, recall that the
symplectic form does have this symmetry, as seen in
Eq.~\eqref{eq:omega-spin-so3}.  To take advantage of this symmetry,
we call $\hat{n}$ a new axis $\hat{z}^{\prime}$, and instead compute
$\frac{1}{2\pi}\oint S_{A}^{z\prime} d\phi_{A}^{\prime}$.  Since each
$\vec{S}_{A}$ rigidly rotates around $\hat{n}$, the integral in one
spin sector will simply be
\begin{align}
  \mathcal{J}_{A}^{\text{spin}} = S_{A}^{z\prime} = \hat{n} \cdot \vec{S}_{A}  \,.
\end{align}

Combining, we see for the generators $J^{2}$ and
$J_{z}$,
\begin{align}
  \mathcal{J} = \hat{n} \cdot (\vec{L} + \vec{S}_{1} + \vec{S}_{2}) = \hat{n} \cdot \vec{J} \,.
\end{align}
Meanwhile, for $L^{2}$, only the orbital sector contributes, and we
have $\mathcal{J} = \hat{n} \cdot \vec{L}$.  This gives us our first
three action integrals,
\begin{align}
  \label{eq:J123}
  \mathcal{J}_{1} &= |\vec{J}| \,, &
  \mathcal{J}_{2} &= J_{z} \,, &
  \mathcal{J}_{3} &= |\vec{L}| \,.
\end{align}

\vspace{.5em}

\subsection{Loop in $R$-$P_R$ space}
\label{sec:loop-r-p_r}

To compute a fourth action variable, we find a loop on the five-torus
(of constant values of the $F_{i}$ mutually-commuting phase-space
functions) in a plane parallel to the $R$-$P_{R}$ plane.  We will
denote the constant values of the $F_{i}$ functions with overbars,
i.e.\ taking the values $H=\bar{\mathcal{E}}, L^{2} =
\bar{\mathcal{L}}^{2}$, and $\overline{L\cdot S_{\text{eff}}}$.
We define $P_{R}$ to be the momentum conjugate to the radial
separation $R$,
\begin{align}
  \label{eq:P_R-def}
  P_{R} \equiv \vec{P} \cdot \hat{R}
  \,.
\end{align}
To show how to construct this loop, we eliminate from the 1.5PN
Hamiltonian all dependence except for $R, P_{R}$, and the values of
constants.  This starts from the definition of
$\vec{L}=\vec{R}\times\vec{P}$, to get
\begin{align}
  L^{2} &= R^{2}P^{2}-(\vec{P}\cdot\vec{R})^{2} \,, \\
  P^{2} &= P_{R}^{2} + \frac{\bar{\mathcal{L}}^{2}}{R^{2}}\,.
\displaybreak[0]
\end{align}
Replacing $P^{2}$ using this relation will eliminate the angular
components of $\vec{P}$ from the 1.5PN Hamiltonian.
To compact the notation, we will again use the scaled variables $r,
p$, with $p_{r}\equiv P_{R}/\mu$, and define the shorthand
\begin{align}
  e_k \equiv \frac{p_r^2}{2} + \frac{\bar{\mathcal{L}}^{2}}{2\mu^2 R^2}
  \,,
\end{align}
which is the Newtonian kinetic energy per reduced mass (and also has
units of $v^{2}$).  Then evaluating the 1.5PN Hamiltonian on this
torus, we find
\begin{widetext}
\begin{align}
  \label{eq:shd30}
  \frac{\bar{\mathcal{E}}}{\mu} =e_k-\frac{1}{r} +  \frac{1}{c^2}
  \bigg\{\frac{1}{2 r^2} -(\nu +3) \frac{e_k}{r} - \frac{\nu \, p_r^2}{2r}
  + \frac{1}{2}(3\nu-1)e_k^2\bigg\}
  + \frac{2G}{c^2\mu R^3}\overline{L\cdot S_{\text{eff}}}
 \,.
\end{align}
This equality demonstrates that we can solve for $P_{R}(R)$ in terms of
$R, \bar{\mathcal{E}}, \bar{\mathcal{L}}$, and $\overline{L\cdot
  S_{\text{eff}}}$ -- thus making a loop while staying tangent to the
torus.  We solve for $P_{R}^{2}$ perturbatively in powers of
$1/c^{2}$, finding
\begin{align}
  \label{eq:P_R-sq-sol}
  P_R^2 = 2 \mu \bar{\mathcal{E}}  +    \frac{ (1-3 \nu )}{c^2} \bar{\mathcal{E}}^2
  + \frac{2GM\mu \left[ \mu + (4-\nu) \frac{\bar{\mathcal{E}}}{c^{2}}  \right]}{R}
  + \frac{
  \left[
    -\bar{\mathcal{L}}^{2} + \frac{(GM\mu)^{2}}{c^{2}}(\nu+6)
  \right]}{R^{2}}
  -  \frac{  \mu G( \bar{\mathcal{L}}^2 + 4\overline{L \cdot S_{\text{eff}}} ) }{ R^3 c^{2}}
  + \mathcal{O}(c^{-4})
  \,.
\end{align}
Here we have collected terms by powers of $R^{-1}$, in anticipation of
performing a Sommerfeld integral, following Damour and
Schäfer~\cite{Damour:1988mr}.  This momentum enters into the
action integral, where the loop is restricted to the $(R,P_{R})$
plane,
\begin{align}
  \mathcal{J}_{4} = \frac{1}{2\pi} \oint P_{R} \ dR
  = \frac{2}{2\pi} \int_{R_{\text{min}}}^{R_{\text{max}}} \left( A + \frac{2 B}{R} + \frac{C}{R^2} + \frac{D}{R^3} \right)^{1/2} \ dR
  \,,
\end{align}
where the coefficients $A, B, C, D$ are \emph{constants} along this loop,
to be read directly from Eq.~\eqref{eq:P_R-sq-sol}.  The factor of 2
comes since the loop runs from one turning point, $R_{\text{min}}$, to
the other, $R_{\text{max}}$, and then back.

To evaluate this integral, we can use the results from Sec.~3 (or
Appendix B) of~\cite{Damour:1988mr}.  The result is in terms of the torus
constants $\bar{\mathcal{E}}, \bar{\mathcal{L}}$, and
$\overline{L\cdot S_{\text{eff}}}$.  We promote these back to their
respective phase-space functions, giving
\begin{align}
  \label{eq:J4}
  \mathcal{J}_4  = -L + \frac{G M \mu^{3/2}}{\sqrt{-2 H}}
  + \frac{GM}{c^2}   \left[  \frac{3 G M \mu^2}{L}
    + \frac{\sqrt{-H} ~\mu^{1/2} (\nu-15)}{\sqrt{32}}
    -\frac{2G\mu^3}{L^3} \SeffdotL \right]
  + \mathcal{O}(c^{-4})
  \,.
\end{align}
\end{widetext}
Unlike the first three actions, the fourth action is not ``exact'' at
1.5PN, but rather we have presented it as a PN series, just like the
radial action in Ref.~\cite{Damour:1988mr}.  This is consistent with
the 1.5PN Hamiltonian itself being a truncated PN series.

The four action integrals we computed are functionally independent, as
can be seen by their different dependence on the original
mutually-commuting phase-space functions $H, J^{2}, J_{z}, L^{2}$, and
$\SeffdotL$.  This corresponds to their loops (all of which are
tangent to a torus) being in linearly-independent homology classes.
The calculations for the fifth action (both as a PN series and
``exact'' at the 1.5PN order) are quite lengthy, so we will present
them in future work.

It is worth noting that at 1.5PN order, spin effects enter the action
integrals, as can be easily seen in Eqs.~\eqref{eq:J123} and
\eqref{eq:J4}.  This is relevant to the method of torus-averaging,
which is used in canonical perturbation
theory~\cite{goldstein2013classical, jose}.  Since the actions depends
on spin, it is easy to see that torus-averaging will differ from
orbit-averaging (over Newtonian orbits) which has been used
extensively in the literature~\cite{Kidder:1995zr, Schnittman:2004vq,
  Racine:2008qv, Kesden:2014sla, Gerosa:2015tea}.
We expect torus-averaging to be more accurate at 1PN and higher orders.

\section{Integrability at 2PN}
\label{sec:integrability_2PN}

The spirit of the post-Newtonian method is perturbation theory in
powers of $1/c$, which opens the door for canonical perturbation
theory applied to Hamiltonian dynamics.  As the KAM theorem
dictates~\cite{goldstein2013classical, jose}, when we add a small
perturbation to an integrable system, and this perturbation breaks
integrability, the perturbed motion is still multiply-periodic and
restricted to $n$-tori, except for resonant tori where
chaos ensues.\footnote{%
The KAM theorem actually gives more precise estimates for the
$\epsilon$ dependence of the chaotic component of phase space;
see Ref.~\cite{jose} for more details.
}

We can take advantage of perturbation theory by treating the 2PN
system as a perturbation upon the 1.5PN Hamiltonian.  We find
deformations to the 1.5PN constants of motion such that the 2PN system
is integrable in the perturbative sense.  This method can be pushed to
higher PN order, but here we only demonstrate it at the first order
where ``exact'' integrability is broken, namely at the 2PN order.
In Sec.~\ref{sec:pert-integr} we explain what we mean by perturbative
integrability, and in Sec.~\ref{sec:method} the method for finding the
deformations to the
constants.  In Sec.~\ref{sec:results} we give the results for the deformed
constants, and discuss some subtle issues in PN integrability in
Sec.~\ref{sec:subtleties}.

\subsection{Perturbative integrability}
\label{sec:pert-integr}

To make the definition of perturbative integrability precise, we will
introduce the ``dominant PN order of'' symbol $[ - ]$.  If a phase-space quantity
is asymptotic to $c^{-2m}$, then it has dominant PN order $m$, i.e.
\begin{align}
  f \sim F(R, P, \chi) c^{-2m} \qquad \longleftrightarrow \qquad
  [f]  \equiv m
  \,,
\end{align}
where $F(R, P, \chi)$ is a $c$-independent phase-space function, and
we employ the $\sim$ symbol of asymptotic analysis~\cite{bender1999advanced}.
The algebra of formal power series tells us how $[ - ]$ interacts
with multiplication, addition, and thus Poisson brackets.
Multiplication is simple,
\begin{align}
  [f\, g] = [f] + [g]
  \,.
\end{align}
When two phase-space
functions have different dominant orders, addition is also simple,
\begin{align}
  [f + g] = \min([f], [g])\, \qquad \text{if } [f] \neq [g]
  \,.
\end{align}
However, if $f\sim -g$, then there will be a cancellation in the
dominant order of $f + g$, and the dominant order of the sum will be
higher than $\text{min}([f],[g])$. Such cancellation can happen in
Poisson brackets, and is necessary for our algebraic definition of
perturbative integrability.

In perturbation theory, equalities only need to be satisfied up to
some sufficiently-small error terms.  Thus for perturbative
integrability, we will replace $\pb{F_{i}, F_{j}} = 0$ with conditions
$\pb{F_{i}, F_{j}} = \mathcal{O}(c^{-2p})$, for some appropriate PN
orders $p$.  If we want perturbative integrability at relative $q$PN
order, we know we want each $\pb{F_{i}, F_{j}}$ to be at least a
factor of $c^{-2(q+1/2)}$ higher than some phase-space quantity, but
what is that quantity?

To answer this question, we define the function $\text{DNC}(f,g)$
which measures what would be the ``expected'' dominant PN order of
$\pb{f,g}$ if there was no cancellation in the leading order
(``dominant non-commutation'').  This expected order has two cases,
corresponding to the leading orders of $f$ and $g$ both contain a
common spin vector or not.  This is because the (inverse) symplectic
form for spins itself carries a power of $S$ and thus $c^{-1}$ [see
Eq.~\eqref{eq:omega-spin-so3}].  Thus we define
\begin{align}
  \text{DNC}(f,g) =
  \begin{cases}
    [f] + [g] - \tfrac{1}{2} \,, & \parbox{10em}{both $f$ and $g$ contain \\ spin at dominant order,} \vspace{0.5em}\\
    [f] + [g] \,, & \text{otherwise.}
  \end{cases}
  \nonumber
\end{align}
If $f$ and $g$ do not have cancellation at the leading order, we see
that $\text{DNC}(f,g)=[\pb{f, g}] $. For example,
$\text{DNC}(R^i,P_i)=0$, but $\text{DNC}(S_{A}^i,S_{A}^j)= 1/2$ for
$i \neq j$.

If $\pb{f,g} = 0$ exactly, then $f$ and $g$ are said to be in involution
up to infinite order. Otherwise we
say that $f$ and $g$ are in involution ``up to $q$PN order'' when
the two equivalent conditions hold,
\begin{subequations}
  \label{eq:pert-int-cond}
  \begin{align}
    \pb{f, g} &\sim \mathcal{O}
    \left(c^{-2(\text{DNC}(f,g)+q +\frac{1}{2} )}\right)  ,  \\
    [\pb{f, g}] &> \text{DNC}(f,g)+q
    \,.
  \end{align}
\end{subequations}
As a consistency check, notice that for the previous examples $(R^i,
P_i)$ and $(S^i_A, S^j_A)$ with $i \neq j$, each pair is not in
involution even at the leading (0PN) order, as would be expected.
Now we define a ``$q$PN constant of motion'' to be a quantity which
is in involution with the $q$PN Hamiltonian up to at least $q$PN
order. Finally, we define $q$PN perturbative integrability in a
$2n$-dimensional phase space when we have $n$ independent phase-space functions
(including the $q$PN Hamiltonian) which are in mutual
involution up to at least $q$PN order.
We will revisit this definition further in Sec.~\ref{sec:subtleties},
and see that it has a shortcoming.

\subsection{Method of finding deformations}
\label{sec:method}

We now construct perturbative constants of motion up to 2PN.  Note
that $J^2$ and $J_z$ always remain exact constants of motion, at
any order, for an SO(3)-invariant Hamiltonian.  Along with the
Hamiltonian, they form a set of three independent mutually commuting
constants of motion. We need to add two more quantities to this list to establish integrability. We
propose that the two required constants of motion are perturbative
deformations of the 1.5PN constants of motion, $L^{2}$ and
$\SeffdotL$, namely
\begin{align}
\widetilde{L^2}   & =      L^2  +  \delta L^2    ,        \label{eq:COM1}   \\
\widetilde{  S_{\text{eff}}\cdot L  }   & = \SeffdotL +    \delta ( \SeffdotL )
\,,
\label{eq:COM2}
\end{align}
where $\delta L^{2}$ and $\delta(\SeffdotL)$ are higher-PN corrections
that we must find.
For every pair, we want involution up to 2PN order [$q=2$ in Eq.~\eqref{eq:pert-int-cond}].
The dominant orders of each of these functions are
$\left[\widetilde{L^2}\right]= [H] = 0$ and $\left[ \widetilde{
    S_{\text{eff} } \cdot L }\right] = \frac{1}{2}$.
Therefore, to satisfy 2PN perturbative integrability, we require
\begin{subequations}
\label{eq:int-conds}
\begin{align}
\big\{    \widetilde{ L^2 } , H    \big\}   & \sim   \mathcal{O}(c^{-5})         \label{eq:aim1}       , \\
\big\{      \widetilde{  S_{\text{eff}}\cdot L } , H    \big\}    & \sim  \mathcal{O}(c^{-6})   , \label{eq:aim2} \\
\big\{      \widetilde{ S_{\text{eff}}\cdot L  } ,   \widetilde{ L^2 }  \big\}  & \sim   \mathcal{O}(c^{-6})   ,       \label{eq:aim3}
\end{align}
\end{subequations}
where $H$ is the 2PN Hamiltonian.

Satisfying these integrability conditions amounts to finding the
deformations $\delta L^{2}$ and $\delta(\SeffdotL)$, which both
proceed following the same approach.  First, the PN orders that are
required to appear in a deformation are identified.  Then we
construct an ansatz for the deformation out of geometrical objects at
these required PN orders, times some coefficients to be determined by
Eqs.~\eqref{eq:int-conds}.  This turns the problem into a systematic
enumerative algebra problem.

At first glance it may seem that this procedure is not systematic, as
there are an infinite number of terms that could appear in such an
ansatz at fixed PN order, but this is not true.  First, the only
quantities that may appear are geometric objects transforming covariantly
under SO(3) rigid rotations:
\begin{itemize}
\item the metric tensor $\delta_{ij}$ (Kronecker delta),
\item Levi-Civita tensor (not the symbol) ${\epsilon}_{ijk}$,
\item the position vector $\vec{R}$, its norm $R$ and unit radial
  vector $\hat{R} \equiv \vec{R}/R $,
\item momentum vector $\vec{P}$, and
\item spin vectors ($\vec{S}_1 $, $\vec{S}_2 $).
\end{itemize}
In practice, it is simpler to construct such ans\"atze from $\hat{R}$
and powers of the scalar $R$, rather than considering $\vec{R}$.
SO(3) covariance requires that these objects automatically
commute with $J^{2}$ and $J_{z}$.  The types of terms allowed in a
deformation have the same tensorial character and parity (scalar,
pseudoscalar, vector, etc.) as the quantity being corrected.
While negative powers $R^{-k}$ can appear in PN expressions, negative
powers of $P$ or $S$ do not.  Now, if we choose a maximum operator
order (number of tensors multiplied together), there are only a finite
number of combinations that can be built at each PN order and operator
order.  Now the problem is indeed enumerative: if a solution is not
found, increase the operator order and try again.

Let us demonstrate by using $\widetilde{L^2}$ as an example.
First, we determine the PN orders necessary for the ansatz of the
deformation $\delta L^{2}$.  Expanding Eq.~\eqref{eq:aim1},
\begin{align}
  \label{eq:COM3}
  \pb{\widetilde{L^2} , H}     ={}& \pb{L^2 ,  H_{2\text{PN}}}  +  \pb{\delta L^2  ,   H_{\text{N}}} \nn\\
  & {}+ \pb{\delta L^2  ,    H_{1\text{PN}}  +  H_{1.5\text{PN}} +  H_{2\text{PN}}}\,, \nn \\
  \sim & {}~ \mathcal{O}(c^{-5})
  \,.
\end{align}
The noncommutation in the first term on the RHS is only with the
spin-spin term, since $L^{2}$ commutes with the orbital part,
\begin{align}
  \pb{L^2 ,  H_{2\text{PN}}} &= \pb{L^2 ,
    H_{\text{SS,2PN}}} \nn \\
  & \sim   \mathcal{O}(S^{2} \, c^{-2}) \sim \mathcal{O}(c^{-4})
  \,.
\end{align}
This is the dominant error that must be cancelled by the terms
involving $\delta L^{2}$, which we see must involve spins.  The
bracket of $\delta L^{2}$ with the Hamiltonian also follows PN
ordering and is dominated by $\pb{\delta L^{2}, H_{\text{N}}}$, with the
other terms being higher-PN.  One must be careful to check what
happens with the spin terms, which potentially reduce PN orders: for
example since $\delta L^{2}$ has spins in its leading order,
$\pb{\delta L^{2}, H_{1.5\text{PN}}}$ is only 1PN order higher than
$\pb{\delta L^{2}, H_{\text{N}}}$, rather than 1.5PN.  Therefore this
condition simplifies to
\begin{align}
  \pb{L^2 ,  H_{\text{SS,2PN}}}  +  \pb{\delta L^2  ,   H_{\text{N}}} = 0    \label{eq:demand_1}
  \,,
\end{align}
with the equality being exact.  To satisfy this, $\delta L^{2}$ will
need to contain two spin in the leading order, which by inspection
must be $\delta L^{2} \sim \mathcal{O}(S^{2} c^{-2})$.

To build an appropriate $\mathcal{O}(S^{2} c^{-2})$ ansatz for
$\delta L^{2}$, we note from Eq.~\eqref{eq:pnc12b} that a factor of $1/c^{2}$ should accompany
either two powers of $\vec{p} \equiv \vec{P}/\mu$, or one power of
$1/R$, and any number of powers of $\hat{R}$.  Since $L^{2}$ is parity
even, we will not use $\epsilon_{ijk}$ to construct the ansatz for
$\delta L^2$: an odd number of $\epsilon$'s makes a parity odd term,
and an even number can be written in terms of $\delta^i_j$.
This yields an ansatz containing terms of the form
\begin{align}
  \label{eq:delta-L-ansatz}
  \delta L^{2} \!\supset\!
  \left(
    \frac{1}{c^{2}}
    \underbrace{S_{A}^{i}S_{B}^{j} P^{k} P^{l} \hat{R}^{m}
      \hat{R}^{n}}_{\text{19 contractions}}
    \,,
    \frac{1}{R c^{2}}
    \underbrace{S_{A}^{i}S_{B}^{j} \hat{R}^{k}
      \hat{R}^{l}}_{\text{6 contractions}}
  \right)
  \,.
\end{align}
Here we mean to take all possible contractions of the two tensorial
forms, where the indices $(A, B)$ label spins in the same way as in
Sec.~\ref{sec:setup}.  This leads to 19 possible contractions
involving two factors of $\vec{P}$, and 6 contractions without
$\vec{P}$, giving us altogether $25$ terms in our most general
ansatz for $\delta L^2$. Since we are taking contractions, the use of
the metric tensor $\delta_{ij}$ is implicit in our ansatz construction.
Our ansatz for $\delta L^2$ then consists of a sum of all these 25 terms with coefficients
to be solved for demanding that Eq.~\eqref{eq:demand_1} be true.

One can employ similar lines of reasoning to construct an ansatz for
$\delta(\vec{S}_{\text{eff}}\cdot\vec{L})$ and solve for the
coefficients so that Eq.~\eqref{eq:aim2}
is satisfied, although it is a more complicated case than
for $\delta L^2$.  Instead of Eq.~\eqref{eq:demand_1}, this time
we demand that Eq.~\eqref{eq:COM103} be satisfied in the next section.
Finally, there may be additional constraints on the terms in the
ans\"atze arising from the requirement that the Poisson bracket
$\pb{\widetilde{L^{2}}, \widetilde{S_{\text{eff}}\cdot L}}$ must also
vanish to the required order, Eq.~\eqref{eq:aim3}.
That is how we finally arrive at the desired
$\widetilde{L^2}$ and
$\widetilde{ S_{\mathrm{eff}} \cdot L }$.
We formed sufficiently general ans\"atze using the
\texttt{AllContractions} and \texttt{MakeAnsatz} commands of the
\textsc{Mathematica} package \textsc{xTras}~\cite{Nutma:2013zea}, which
works in the \textsc{xAct/xTensor} suite~\cite{JMM:xAct,
  MARTINGARCIA2008597}.  Our result may be verified by the
\textsc{Mathematica} notebook which accompanies this
article~\cite{MmaSupplement}.

\subsection{The deformed constants}
\label{sec:results}
Following the above procedure to find a deformation to $L^{2}$, we
write this deformation as
\begin{align}
  \widetilde{L^2} = \underbrace{ L^2}_{\mathrm{0PN}}   +   \underbrace{ \delta L^2}_{2\mathrm{PN}}
  \,.
  \label{eq:COM1000}
\end{align}
For brevity we will define the symmetric tensor
\begin{align}
  \label{eq:hTensDef}
  h^{ij} \equiv \frac{p^{i}p^{j}}{2} - \frac{r^{i}r^{j}}{r^{3}}             
  \,,
\end{align}
where we again used the scaled variables $\vec{p}\equiv\vec{P}/\mu$,
$r=R/GM$.  Notice that $h^{ij}$ has units of $v^{2}$, and that
the trace is
\begin{align}
  h \equiv \delta_{ij}h^{ij} = H_{N}/\mu                  \label{eq:h_scalar}
  \,.
\end{align}
Then we can write our deformation as
\begin{align}
  \label{eq:deltaL2}
  \delta L^{2} = \frac{-2\nu}{c^{2}}
  \bigg[  &
   \frac{m_{2}}{m_{1}}S_{1}^{i}S_{1}^{j}  h_{ij}
  + S_{1}^{i}S_{2}^{j}  \left(    h_{ij} - \delta_{ij} \frac{h}{2}    \right)  \\
   &   + (1 \leftrightarrow 2)
  \bigg]
  \,. \nonumber
\end{align}
We are also free to add
arbitrary constants times $S_{1}^{2} h/c^{2}$ and $S_{2}^{2}
h/c^{2}$ without affecting integrability.

Proceeding similarly for $\SeffdotL$, we decompose the
deformation as
\begin{align}
  \label{eq:tildeSeffdotLDecomp}
  \widetilde{S_{\text{eff}}\cdot L} = \underbrace{\SeffdotL}_{\text{0PN}}
  + \underbrace{\delta_{1}(\SeffdotL)}_{\text{0.5PN}}
  + \underbrace{\delta_{2}(\SeffdotL)}_{\text{1.5PN}}
  \,.
\end{align}
The two deformations are
\begin{align}
  \label{eq:delta1SeffdotL}
  \delta_{1}(\SeffdotL) = {}&\frac{1}{4} \vec{S}_{1}\cdot \vec{S}_{2}
  \,,
\end{align}
\begin{widetext}
\begin{align}
\label{eq:delta2SeffdotL}
  \delta_{2}(\SeffdotL) = {}& \frac{1}{c^{2}}
  \bigg[
    \sigma_{1} \frac{m_{2}^{2}}{M^{2}} S_{1}^{i} S_{1}^{j} h_{ij}
    +
    \frac{1}{8}(3+2\nu) S_{1}^{i} S_{2}^{j}  h_{ij} %
    + (1 \leftrightarrow 2)
  \bigg] \,.
\end{align}
We are also free to add arbitrary constants times
$S_{1}^{2} h/c^{2}, S_{2}^{2} h/c^{2}$, and
$(\vec{S}_{1}\cdot \vec{S}_{2}) h/c^{2}$ without affecting
integrability.
The cancellations happen as
\begin{align}
  \underbrace{  \pb{   \SeffdotL  , H_{\text{SS,2PN}} }  }_{\text{\parbox{12em}{both orbital and spin PBs; $\mathcal{O}(c^{-4})$ and $\mathcal{O}(c^{-5})$}}}
  + \underbrace{    \pb{   \delta_1 ( \SeffdotL ), H_{1.5\text{PN}}  }   }_{\text{\parbox{4.8em}{spin PBs; $\mathcal{O}(c^{-4})$}}}
  + \underbrace{    \pb{    \delta_1 ( \SeffdotL ) , H_{\text{SS,2PN}} }   }_{\text{\parbox{4.8em}{spin PBs; $\mathcal{O}(c^{-5})$}}}
  + \underbrace{    \pb{   \delta_2 ( \SeffdotL ) , H_{\text{N}} }   }_{\text{\parbox{6em}{orbital PBs; $\mathcal{O}(c^{-4})$}}}    =0   \label{eq:COM103}
  \,,
\end{align}
\end{widetext}
with the equality being exact, where $H_{\text{SS,2PN}}$ is defined
in Eq.~\eqref{eq:int1e}.  Below every
Poisson bracket, we indicate both the PN orders arising, and what kind
of PBs (orbital or spin) are needed to expand each term.
With these corrections, we have fulfilled the required level of
commutation given in Eqs.~\eqref{eq:int-conds}.  In fact, we
slightly exceeded this goal, achieving
\begin{subequations}
\label{eq:int-achieved}
\begin{align}
\label{eq:int-achieve-L2-H}
\big\{    \widetilde{ L^2 } , H    \big\}   & \sim   \mathcal{O}(c^{-6}) \,, \\
\label{eq:int-achieve-SeffL-H}
\big\{      \widetilde{  S_{\mathrm{eff}} \cdot L } , H    \big\}    & \sim  \mathcal{O}(c^{-6})  \,, \\
\label{eq:int-achieve-SeffL-L2}
\big\{      \widetilde{  S_{\mathrm{eff}} \cdot L } ,   \widetilde{ L^2 }  \big\}  & \sim   \mathcal{O}(c^{-7}) \,.
\end{align}
\end{subequations}
Therefore, along with the 2PN Hamiltonian $H, J^2$, and $J_z$, the
deformed constants $\widetilde{L^{2}}$ and
$\widetilde{S_{\text{eff}}\cdot L}$ now form a set of 5 independent,
mutually commuting constants of motion at 2PN order, thereby
establishing the integrable nature of the BBH system at this
order.

It is worth comparing our results to the widely-used results based on
orbit-averaging (over a Newtonian orbit)~\cite{Kidder:1995zr,
  Schnittman:2004vq, Racine:2008qv, Kesden:2014sla, Gerosa:2015tea}.
Racine found~\cite{Racine:2008qv} that the combination
$\vec{S}_0 \cdot \vec{L}$ is conserved by what we call
$\avg{d/dt}_{N}$, the Newtonian-orbit average of the 2PN EOMs.  Here
$\vec{S}_0 \equiv (1+m_2/m_1) \vec{S}_1 + (1+m_1/m_2) \vec{S}_2$ was
introduced by Damour~\cite{Damour:2001tu}. Two comments are in order.
First, $\vec{S}_0 \cdot \vec{L}$ differs at its leading order from
$\SeffdotL$ and therefore $\widetilde{S_{\eff}\cdot L}$.  Since spins
and $\vec{L}$ are all constants at Newtonian order, applying the
Newtonian-orbit-average to form $\avg{\SeffdotL}_{N} =\SeffdotL$ does
not recover $\vec{S}_{0}\cdot \vec{L}$.  Second, as mentioned at the
end of Sec.~\ref{sec:act_1.5PN}, a more accurate average is not
over the Newtonian orbit, but on the phase-space torus formed by
level sets of the five constants of motion.  The torus-average will
already differ at 1PN order from the Newtonian-orbit average.
We can confirm using the 2PN Hamiltonian and averaging
over the Newtonian orbit the two independent equalities,
\begin{align}
  \avg{\frac{d}{dt}}_{N} \vec{S}_{0}\cdot \vec{L} ={}& 0 \,,&
  \avg{\frac{d}{dt} \vec{S}_{0}\cdot \vec{L}}_{N} ={}& 0 \,.
\end{align}
However, we should expect that the torus-average will differ.  More precisely, with
the 2PN Hamiltonian and no averaging,
\begin{align}
  \frac{d}{dt} \vec{S}_{0}\cdot \vec{L} = \mathcal{O}(S^{2} c^{-2}) = \mathcal{O}(c^{-4})
  \,.
\end{align}
Thus while Newtonian-orbit averaging gives a cancellation of this leading order, we
expect the more accurate torus average to be nonzero at the order
\begin{align}
  \avg{\frac{d}{dt} \vec{S}_{0}\cdot \vec{L}}_{T} = \mathcal{O}(c^{-6})
  \,.
\end{align}
Notice this is the same level of conservation that we achieved in
Eq.~\eqref{eq:int-achieve-SeffL-H}, but our result is valid
instantaneously, that is, without resorting to averaging.

\subsection{PN constancy and integrability revisited}        \label{sec:subtleties}

Our algebraic definition of PN involution and integrability introduced
in Sec.~\ref{sec:pert-integr} has a shortcoming.
To understand this, let's examine the timescales on which phase-space
quantities vary.  For some quantity $f$, when evolved with the full
$n$PN Hamiltonian $H^{n\text{PN}}$ (not the $n$PN contribution to the
Hamiltonian), we can approximate the timescale of variation with
\begin{align}
  \label{eq:time-def}
  T_n(f)   \equiv \frac{f}{\pb{f, H^{n\text{PN}}}}
  \,.
\end{align}
For example, the orbital (or Newtonian) timescale is
\begin{align}
  T_{N}   \equiv T_{0}(R^{i})  \approx \sqrt{\frac{R^{3}}{GM}}
  \,.
\end{align}
Now, with the algebraic definition of PN integrability given in Sec.~\ref{sec:pert-integr},
$\SeffdotL$ is a 1.5PN constant of motion.
But let us examine the timescale of its variation, in units of the
orbital time. We cannot use $H^{1.5\text{PN}}$ for this, since
$\SeffdotL$ and $H^{1.5\text{PN}}$ commute.  The
timescale of variation is controlled by the 2PN Hamiltonian, and one
can check
\begin{align}
  T_{2}(\SeffdotL) \sim \mathcal{O}
  \left(\left(\frac{v}{c} \right)^{-3}
    T_{N} \right)
  \,,
  \label{eq:comparison1}
\end{align}
implying that $\vec{S}_\text{eff} \cdot \vec{L}$ varies on a timescale
that is only 1.5PN longer than $T_{N}$, rather than the expected 2PN
orders longer.  Therefore, $\SeffdotL$ is not a
1.5PN constant from the criterion of comparing timescales,
and the BBH system cannot yet be called integrable at 1.5PN order despite the
existence of five exactly commuting constants at this order.

The key point is that $H_{n\text{PN}}$ may sometimes induce
variations in a quantity $f$ at a timescale which is only $(n-1/2)$PN
orders larger than $T_{N}$, rather than $n$PN orders larger.  As was
emphasized in Secs.~\ref{sec:SecPNC}, \ref{sec:Hamilton}, and
\ref{sec:pert-integr}, this happens because of the factor of $c^{-1}$
in spin, and the form of the spin Poisson bracket.
Therefore, to establish if a quantity is a constant of motion on an
$n$PN timescale will generally involve examining the
$(n+1/2)$PN Hamiltonian.

To conservatively satisfy the timescale analysis, we revise the
earlier definition of $q$PN constancy and integrability by using the next
order, $(q+1/2)$PN, Hamiltonian, instead of the $q$PN Hamiltonian.
However, we only introduce relative $q$PN corrections to our deformed
constants.  We have checked that the five quantities
$H, J_{z}, J^2, \widetilde{ L^2 }$ and
$\widetilde{S_{\text{eff}}\cdot L}$ ($H$ now being the 2.5PN
Hamiltonian~\cite{Steinhoff:2010zz}) are also in mutual involution up
to 2PN according to our revised definition, even though the last two
quantities were derived in Sec.~\ref{sec:method} by only considering the 2PN
Hamiltonian.
This calculation is also verified in the supplement to this
article~\cite{MmaSupplement}.
In terms of timescales, we now satisfy
\begin{align}
    T_{2.5}\left({\widetilde{S_{\eff}\cdot L}}\right) \sim \mathcal{O}
  \left(\left(\frac{v}{c} \right)^{-5}
    T_{N} \right)
  \,.
\end{align}
Hence, we have established the integrable nature of the
BBH system at one PN order higher (2PN) than what was earlier known
(1PN) on the basis of timescale of variation.

\section{Discussion}
\label{sec:discussion}

In this paper, we studied the problem of integrability at two levels:
1.5PN and 2PN. At 1.5PN, where exact integrability had already been
known~\cite{Damour:2001tu}, we evaluated four (out of five) action
variables, with the fourth one being a perturbative PN series.
At 2PN order, by adding corrections to the 1.5PN mutually
commuting constants of motion, we constructed 2PN perturbatively
commuting quantities. This proves the integrable nature of the BBH
system at 2PN in a perturbative sense. Our construction required us to
propose appropriate definitions of PN involution and integrability.
Proving perturbative integrability at 2PN and higher is more delicate
than at 1.5PN, since the 1.5PN commutation does not require
perturbation theory. We presented a systematic method to find
higher-PN corrections to mutually commuting constants of motion,
forming an ansatz by enumerating possible tensor expressions, turning
the problem into linear algebra. We therefore expect our method to be
useful in extending integrability to even higher PN orders.

By now a large number of authors have studied the problems of
integrability or chaos in the BBH system in post-Newtonian theory,
either numerically or analytically.
Importantly, while Hartl and Buonanno~\cite{Hartl:2004xr} did find chaos
in the PN BBH system, they found it is only present in a small
component of phase space.
The constants of motion we have constructed apply to the invariant
tori in the non-chaotic regions of phase space, i.e.\ the majority of
the volume.  This improves the outlook for using
perturbative integrability as a tool for generating highly-accurate
and efficient waveform models.

To employ integrability for efficient waveform modeling, the current
work will have to be extended in a number of natural ways.
We plan to present the exact (at 1.5PN) fifth action variable and its
PN expansion in a future article, also yielding all the frequencies
of the system in closed form. Work still needs to be done towards finding the angle variables.
These action-angle variables are related to the recent
Keplerian-like solution for the eccentric, spinning BBH system at
1.5PN~\cite{Cho:2019brd}. These action-angle variables can be pushed
to 2PN order and beyond via perturbation
methods. This will fail for the small
chaotic region of phase space, and more care will be needed near
resonances.

This opens the possibility to construct an analytic waveform model for
the completely generic system, without needing to e.g.\
orbit-average~\cite{Kidder:1995zr, Schnittman:2004vq, 
  Racine:2008qv}, precession-averaging~\cite{Kesden:2014sla,
  Gerosa:2015tea, Khan:2018fmp, Pratten:2020ceb}, or expand in
powers of eccentricity~\cite{Klein:2010ti, Klein:2018ybm}.
As discussed at the end of Sec.~\ref{sec:results}, we expect the
time derivatives of the orbit-averaged constants to have errors at
relative 2.5PN order, when averaged over the true orbits, rather than
over Newtonian orbits.  This is the same level of error in the time
derivatives of our instantaneous constants, i.e.\ without
needing to average.
We hope to see our integrability results applied to future analytical
waveform models such as the \texttt{Phenom} family.

A difficulty will
arise at 2.5PN order, where the dynamics are no longer conservative.
Starting at this order, the ``constants'' of motion will now vary with
time. One possible approach will be the formalism of non-conservative
classical dynamics~\cite{Galley:2012hx, Galley:2014wla,
  Galley:2016zee, Tsang:2015cua}, which has a Hamiltonian
version. Even if the non-conservative approach proves difficult, the
conservative sector of the dynamics can still be pushed to higher PN
order, and the time-evolution of the ``constants'' imposed afterwards
through order reduction.
\vspace{1.5em}

\acknowledgments

We would like to thank
Samuel Lisi
and
Clifford Will
for helpful discussions, and Davide Gerosa for the initial motivation
to investigate post-Newtonian spin dynamics that eventually led to
this work. The work of JG was
partially supported by the Natural Sciences and Engineering Research Council of 
Canada (NSERC), funding reference \#CITA 490888-16, \#RGPIN-2019-07306.

\bibliography{pn}

\end{document}